# Reverse analysis of a spiral injector to find geometrical parameters and flow conditions using a GA-based program

Maziar Shafaee[1*], Armin Abdehkakha[2], Abbas Elkaie[3]

[1] Professor Assistant, *Faculty of New Sciences and Technologies, University of Tehran, Tehran, 143995-5941, Iran*
[2, 3] Master's Student, *Faculty of New Sciences and Technologies, University of Tehran, Tehran, 143995-5941, Iran*

**Abstract**     In this article, the effect of geometrical parameters and flow conditions on the performance of a swirl atomizer is studied. Dimensional analysis and experimental investigations are utilized to define significant terms. The PDA system used for the measurements was able to supply information about the size, concentration, and particle velocity at each measurement location. The orifice diameter, the spiral cone angle and also the flow Reynolds number, which is defined based on the injector orifice diameter, plays an important role in spray quality, and their significance is summarized in a correlation. In order to achieve the appropriate combination of design variables that satisfy the design constraints, a GA-based program was used in a reverse analysis process. Finally, the advantages of human inspection were employed to provide true best performers from a small group of final answers.

______________________________________________________________________________________________

## 1. Introduction

Atomizers are used in many industrial applications to atomize different flows to droplets [1]. In order for the atomizers to have a high performance, the drop size distribution must have certain characteristics (small, and good drop size distribution). Physical properties and geometrical parameters have an influence on the spray quality and we need to understand these effects to achieve an optimum atomization [1-3].

Swirl atomizers accelerate the liquid into a swirl chamber. As a result, a centrifugal force develops and reshapes the fluid into a hollow cone which further disintegrates into ligaments and drops. Hollow and full cones are two common types that are produced in this type of atomizer. The air core of a hollow cone assists in ligament breakup and produces smaller droplets. These atomizers were used for propulsion systems, chemical reactors, gas cooling systems, and dust control facilities. The swirling motion of the atomizer is the result of the spiral passage (groove) inside the injector which not only develops the motion but also controls the distribution of the fluid in the final spray [4, 5].

Morad et al. studied the spray characteristics of a liquid-liquid coaxial swirl atomizer at different mass flow rates. They measured the Sauter mean diameter, and droplet axial and radial velocities using a system of PDA. Their study showed that smaller particles possess a smaller magnitude of velocity in contrast to larger particles which have a greater magnitude of velocity. Their study also revealed a peak in the velocity magnitude of particles in the radial coordinates of the spray [6].

In another study by Radke on a liquid–liquid swirl injector, further insight into geometric and flow parameters has been provided. The output diameter of an atomizer and flow properties have significant effects on sheet and ligament breakup.

Corresponding author. Tel.: +989190110200
*E-mail addresses*: mshafaee@ut.ac.ir (M. Shafaee), armin.abdeh@yahoo.com (A. Abdehkakha), a.elkaie@ut.ac.ir (A. Elkaie)

He supposed that increasing the Re effect on SMD is similar to increasing the output diameter [7].

Ibrahim et al. delved further into instability mechanisms and the breakup of liquid sheets. Increasing viscosity reduces radial and tangential velocity components, and consequently the cone angle of a spray [8].

There are different experimental methods to define the Sauter mean diameter and study the spray characteristics. Optical drop sizing methods and image processing techniques are very popular and often desired. The complexity of the atomization process makes the prediction difficult and the results are still reported in terms of various correlations. Shafaee et al. [9] used an image processing technique to develop a correlation between the spray cone angle and the weber number. By using other methods, Merington and Richardson [10] proposed a correlation for the Sauter mean diameter of spray produced by a circular orifice atomizer.

Despite the relatively complex structure of a swirl atomizer, it is quite well known and has its own specific applications, particularly in the field of combustion. The wide particle distribution generated mainly over the hollow air core decreases the rich mixture zone which results in a combustion performance decrement and a soot production increment. In this article, we are mainly concerned with geometrical parameters and flow properties.

The SMD has a significant impact on the performance of jet engines. A small SMD spray facilitates more rapid evaporation, and therefore can lead to a more efficient combustion and fuel consumption. However, a large SMD spray may lead to improper combustion and therefore can decrease the system performance and thrust. Injector design and operational conditions should be set according to the level of mixing which can be represented through the SMD value. This paper is mainly concerned with defining the operational and geometrical parameters to produce a qualified spray with predefined SMD and also with practical considerations.

The process of design with only a limited amount of available data is an important challenge in engineering. The design procedures will not only be difficult but also may not lead to an optimal solution due to the lack of sufficient data. Conventional techniques, such as applying trial and error in design process, lead to frequent iterations. These methods may resolve the problem eventually, but there are several other unexamined possible solutions. Furthermore, these methods are extremely time-consuming. The proposed solution is to use optimization techniques. Optimization techniques are significantly efficient in comparison with trial and error-based methods [11, 12].

Genetic algorithms are one of the optimization techniques and the high efficiency of GA has been proven in a wide range of optimization problems. For a certain limited set of initial data, it was revealed that the GA method is more capable of establishing the remaining unknown design parameters compared with the trial and error method [11, 13].

In a reverse analysis process, some of the final goals such as geometrical and operational limitations, as well as some of the design variables, are known. In this case, consistent solutions with known inputs are produced using GA. Hartfield et al. have performed extensive research studies on the performance of GA in a number of reverse analysis systems. These results show that GA is able to achieve good results, however, it may require a human inspection as well to produce the truly best performance from a small group of final solutions. This method would be able to provide the optimal design parameters which are not realizable in the trial and error approach [14].

The method which is proposed in this paper is to use a GA-based program to solve the problem. In this method, the unknown variables are calculated based on design objectives, first defining the definite variables and objectives. The advantages of this method include not only a faster solution than a trial and error solution, but also provide a large set of acceptable solutions. In this paper, different designs of injectors may satisfy our main criterion. Thus, before the final design is drafted a thorough analysis of design space is essential, which is implemented by the GA-based approach.

## 2. Experimental setup

The swirl atomizer and its geometrical parameters, including the spiral angle, and the length and diameter of the orifice, are sketched in Fig. 1. Table 1 summarizes the geometrical parameters of the atomizer.

Table 1  Geometrical parameters values of atomizer

| Parameter | Value |
|---|---|
| $2\beta$ | 90° |
| $\alpha$ | 21° |
| $L_c$ | 6 (mm) |
| $d_c$ | 3 (mm) |
| $2\theta_1$ | 56° |
| $2\theta_2$ | 154° |

Where $\beta$ and $\alpha$ are the cone angle and the spiral angle, respectively; $L_c$ is the length of the atomizer; $d_c$ is the orifice diameter and $\theta_1$ and $\theta_2$ are the nozzle convergence angles.

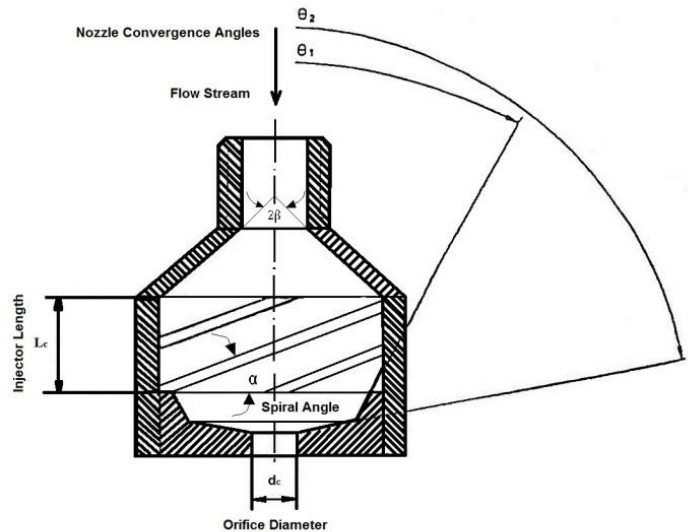

Fig. 1  Sketch of spiral atomizer

A Phase Doppler Anemometer (PDA) was employed to measure the velocity and diameter of droplets. Measurements are made based on phase Doppler interferometric theory, whereby light from two incident laser beams are scattered by particles entering the measurement volume, defined by the intersection of the two beams. In this study, the PDA system consists of an Ar-ion laser, optical fibers, a transmitter, a stroboscope, a signal processor, an oscilloscope, a digital camera, a fixture, a traversing system, and a computer for performing analyses.

Injectors are located inside a fixture and the fixture is placed on a motor with a three-dimensional traversing system. The traversing system was controlled with a computer code automatically. The measurement points are arranged in circular arrays of three different radiuses in addition to a centre point. In each radius, there are eight measurement points located at fixed angle positions with steps of 45 degrees. The arrangement of the measurement points available for the PDA system is shown in Fig. 2. The test liquid was water which was provided from a pressurized reservoir. A pressurized tank of nitrogen was provided with an adjustable pressure control on the liquid tank. the sketch of the apparatus configuration is shown in Fig. 3.

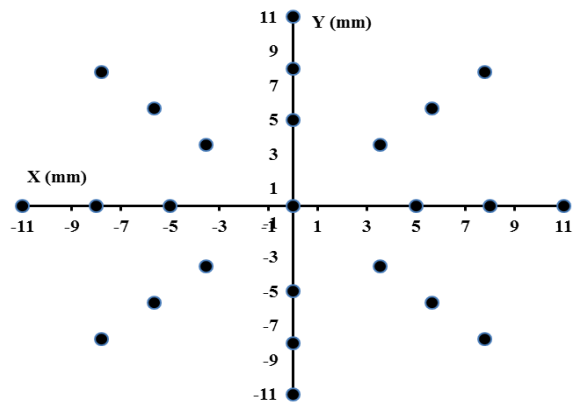

**Fig. 2** Arrangement of measurement points on the spray screen

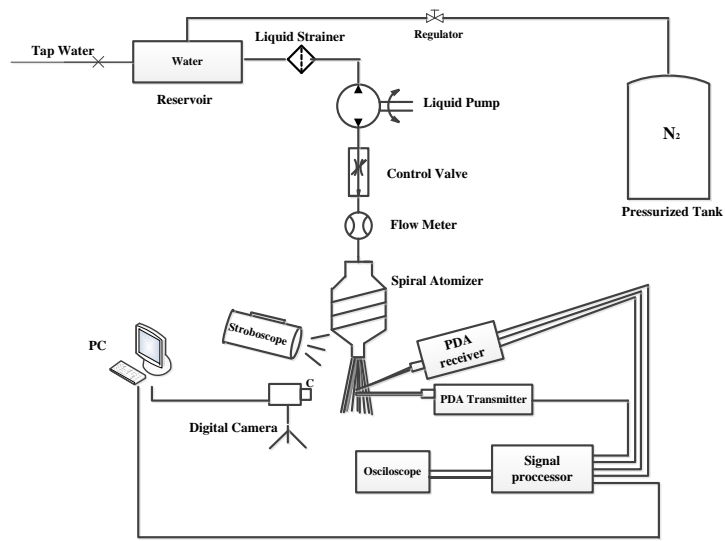

**Fig. 3** Sketch of experimental setup

These measurement points are located on the horizontal plane at a distance of 30 millimeters from the orifice of the atomizer. The order of measurement was determined as being from the centre point toward the points on the larger circles in a counterclockwise sequence. There are 25 points of measurement in total.

**3. Results**

*3.1. Dimensional analysis*

Dimensional analysis can be a powerful method for investigating the unknown physical relations between dimensionless parameters. The first step in a dimensional analysis concerns defining all the significant physical parameters. Here, we are interested in SMD and the possible relation could be as follows:

$$SMD = f\left(\rho_L, \mu_L, U_L, L_c, d_c, \sin\alpha\right) \tag{1}$$

Where $SMD$ is the Sauter mean diameter; $\rho_L$ is the density of the liquid; $\mu_L$ is the viscosity of the liquid; $U_L$ is the axial velocity; $L_c$ is the length of the atomizer; $d_c$ is the orifice diameter and $\alpha$ is the spiral angle.

A standard dimensional analysis (Buckingham's π-theorem) shows:

$$\frac{SMD}{d_c} = f\left(Re, \frac{L_c}{d_c}, \sin\alpha\right) \tag{2}$$

Where, $Re$ is the Reynolds number and would be defined here as $Re = \frac{\rho_L U_L d_c}{\mu}$.

The ratio of the spiral chamber length to the orifice diameter and the cone angle are the most important geometrical attributes of the atomizer. The Reynolds number represents the importance of the fluid properties, i.e., viscosity and density. The characteristic flow speed in the Reynolds is the result of the particular geometry of the atomizer.

Fig. 4 shows the variation of concentration along the radius of the spray. The concentration is almost constant in circular regions. The measurement was carried out by using PDA and in order to improve the accuracy, averaging is applied to the measurement points arranged in the inner, middle and outer circles.

Fig. 4a shows the contour of the droplet concentration. The colour red indicates a high concentration of droplets in the central region of the cone-shaped spray. Fig. 4b illustrates the 3D schematic of the size distribution and Fig. 4c shows the concentration variation along the radius.

*3.2. Study of the effect of orifice diameters*

To investigate the effect of the orifice diameter on the characteristics of atomization, four atomizers with orifice diameters of 2.5, 3, 3.5, and 4 mm were used. PDA measurements showed that a decrease of the orifice diameter causes the axial and radial components of the velocity to rise, but on the other hand the tangential component would decrease. The tangential component is very small compared to the axial and radial components. As a result, the magnitude of the velocity is reduced (See Eq. (3)). Fig. 5 shows the changes of velocity along the radius for the different orifice diameters. An increase in the orifice diameter reduces the overall speed of the particles, but asymptotically loses its effectiveness.

$$A = \sqrt{V_r^2 + V_t^2 + U_L^2} \tag{3}$$

In Eq. (3), $V_{tot}$ is the total velocity whereas $V_r$ and $V_t$ are the radial and tangential components of the velocity respectively.

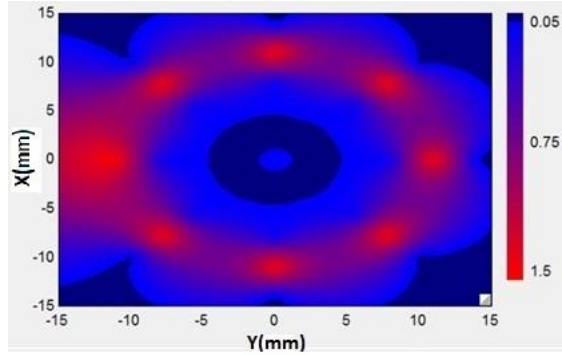

(a)

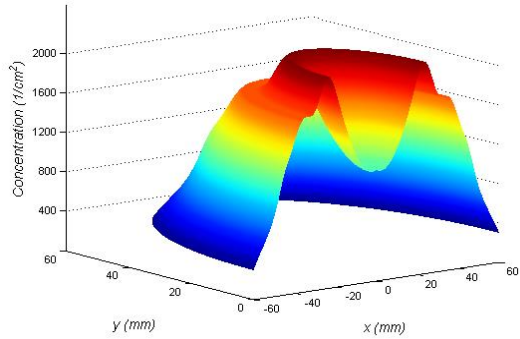

(b)

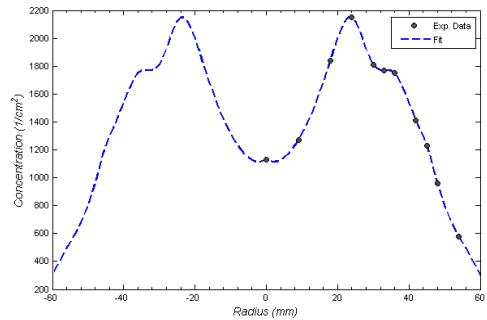

(c)

**Fig. 4** The variation of concentration along the radius of spray

Ignoring the frictional losses, a large amount of static pressure turns into dynamic pressure, as the flow passage narrows and, therefore, the flow speed increases. The atomizer with a smaller orifice diameter produces a faster atomization regime. The results of the experiment confirm this fact. Fig. 6 shows that the variation in the ratio of radial to axial velocity compo-nents is well affected by the radius. According to the figure, this proportion that represents spray cone angle decreases with the increase of the orifice diameter.

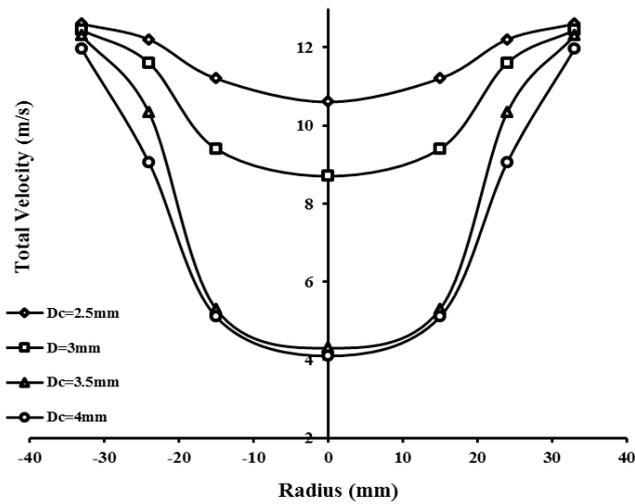

**Fig. 5**　The changes of total velocity along the radius of spray for different orifice diameters

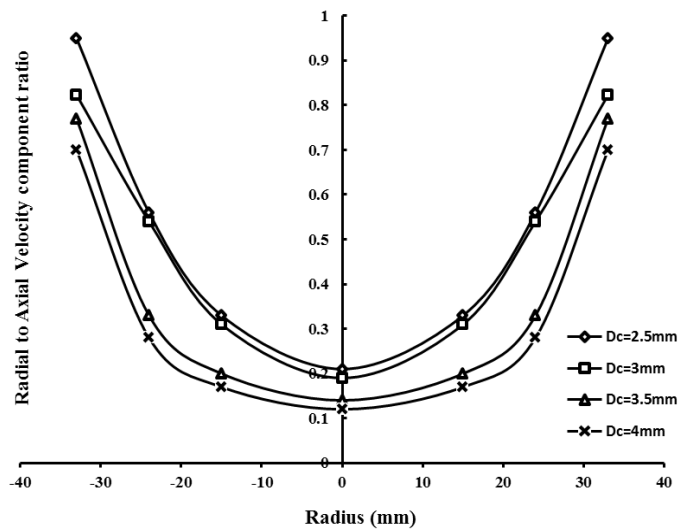

**Fig. 6**　Variation in ratio of radial to axial velocity components for different orifice diameters

In fact, the broad range of atomization applications demands different particle characteristics. SMD is widely used as a characteristic diameter to compare atomization quality. Fig. 7 show changes of SMD based in changes of the orifice diameter. It is observed that decreasing the orifice diameter reduces SMD and therefore could lead to further increases in spray quality.

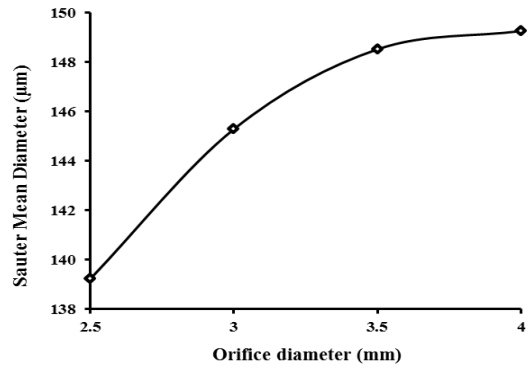

**Fig. 7**    Variation of SMD along the orifice diameter

*3.3. Study of the spiral angle effect*

Spiral angle has an effect on spray quality. Three atomizers with different spiral angles of 15, 25 and 35 degrees were used to investigate this effect. The measurements have been performed on 11 circular point arrangements at the horizontal plane at a distance of three centimetres from the orifice. The magnitude of radial and axial velocity components increases with the increasing radius. The tangential velocity component is almost constant and its value is very small compared to other components at all the measurement points, so this component does not play a significant role in the total magnitude of the velocity and therefore would be negligible as a consequence. Fig. 8 shows the variation of absolute velocity with respect to the radius for three spiral angles. On the boundaries, the speed is about 16 m/s and decreases to 6.5 m/s towards the centre of the spray. In the atomizer with a greater spiral angle, the radial velocity is smaller because the centrifugal force would be lower. In pressure swirl atomizers, the axial velocity and, therefore, the velocity magnitude of the spray initially increases with increasing radial distance, and after reaching a maximum, it decreases. The slowdown in the outer region of the spray cone is mainly due to the vortices created there. Das [15] numerically showed that two vortices that develop from the cone of the spray are amplified by a further increment of the spray angle. This is related to the fact that by increasing the spray angle, the interaction of the vortices decreases.

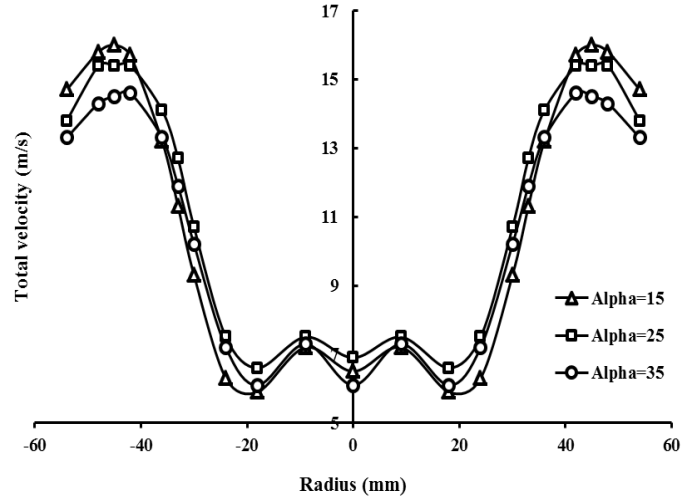

**Fig. 8** Variation of absolute velocity with respect to radius for three spiral angle

Decreasing the spiral angle at a constant length of the spiral chamber is equivalent to increasing the number of grooves and therefore increasing the centrifugal force that is exerted on the fluid through the chamber. As a result, the growth of the radial velocity is greater than the axial velocity component. However; this change is very small and negligible. Fig. 9 shows how SMD changes with respect to the spiral angle. According to the results, the spiral angle is inversely proportional to SMD. This fact occurs because the centrifugal force is reduced.

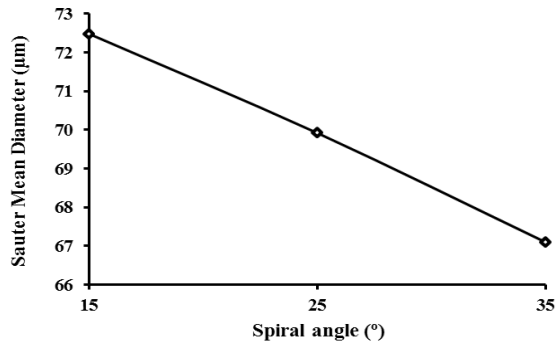

**Fig. 9** SMD changes with respect to spiral angle variation

*3.4. Study of the spiral chamber-length on hydrodynamics*

Two atomizers with the same operating conditions and the different chamber lengths of 3 and 6 mm were tested. Reducing the length of the atomizer energizes the flow by creating more centrifugal force and increasing the tendency of the flow to move in the axial direction. Thus, the particles produced in the chamber of a shorter length are faster and more penetrable as a result of the conversion of the axial velocity component into a radial velocity component by the presence of more centrifugal force. For the atomizer that is 6 mm in length, the radial velocity component possesses a greater magnitude away from the centre toward the outer zone, and this is well confirmed by experiments. The short length of the chamber causes the spray to concentrate in the central zone but by increasing the length, the concentration and momentum of the outer zone increases. The reason for this could be the change in the radial component of velocity.

Fig. 10 shows the variation of SMD in terms of the atomizer's length. According to this plot, SMD increases with increasing length. The atomizer's length is proportional to the spiral chamber length and by increasing the length of the chamber, axial velocity is reduced. The radial velocity, however, is increased due to the increasing influence of the centrifugal force. Friction reduction on the fluid surface reduces the SMD as well.

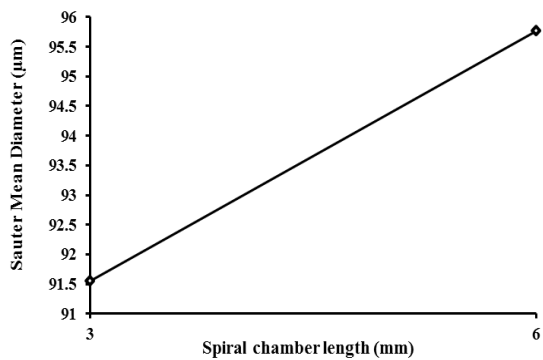

**Fig. 10** SMD changes with respect to spiral chamber length variation

*3.5 Study of the Reynolds number effect*

The Reynolds number is a dimensionless number that takes into account the parameters such as characteristic length, speed, and flow properties and has a considerable effect on the atomizer's performance and SMD. In our test, the only variable parameter in the Reynolds number is the characteristic length (i.e., orifice diameter; $d_c$). Fig. 11 shows the variation of SMD versus the Reynolds number. The Sauter mean diameter of the droplets increases with an increase in the Reynolds number. For higher values of the Reynolds numbers, there is no significant effect on SMD.

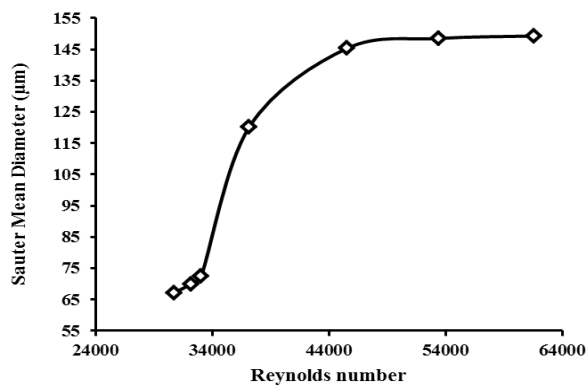

**Fig. 11** Variation of SMD according to Reynolds number

## 4. Discussion

An SMD correlation is presented in Eq. (4) that is based on the flow conditions and geometric parameters. The length of the atomizer has little impact on SMD therefore the correlation is only a function of Re, Sinα and, $d_c$.

$$\frac{SMD}{d} = -404.41\left(d_c^{-0.849}\right) + 113.034\left(\sin^{-0.095}\alpha\right) - 0.000132(Re^{1.271}) + 196.18 \tag{4}$$

Fig. 12 shows the comparison of the experimental and correlation results. The suggested correlation has an error percentage of less than 8%.

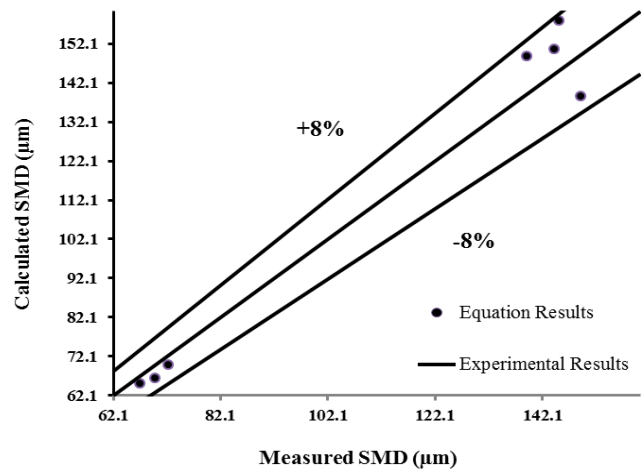

**Fig. 12** Comparison of experimental and correlation results

As was mentioned, the goal is to produce droplets within a predefined range of SMD. This is achieved by using the correlation equation (Equation 4) for SMD and the concept of reverse analysis which was introduced earlier in this paper.

The correlation consists of two geometric variables ($\alpha$ and $d_c$), a flow parameter (*Re*) and finally an operational variable. In order to implement a method based on GA, the geometric and flow parameters are taken as design variables and the operational parameters of SMD is taken as the objective function. The reverse analysis process requires a certain range of SMD to be considered. For example, in the case that SMD is equal to 40μm the set of solutions associated with flow conditions and geometrical parameters are calculated. Then, by considering implementation constraints and possibility criteria, the optimal set of solutions are obtained. The results of the aforementioned example are presented in Fig. 13.

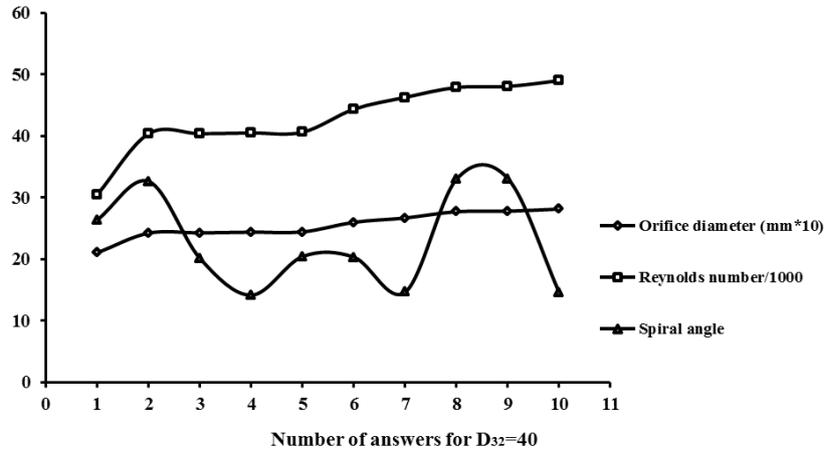

**Fig. 13**   Results of GA algorithm for $D_{32}=40$

In Fig. 13, the effect of the Reynolds number and two other geometrical parameters are plotted and for a particular application a specific solution may be selected. For example, for low-thrust propulsion systems which require a small Reynolds number, the set of solutions located on the left side of Fig. 13 are available. Depending on manufacturing capabilities and layout space, a smaller or larger $d_c$ might be used.

## 5. Conclusion

The effect of the geometric parameters on atomizer performance is examined and the following two significant conclusions are drawn:
1. The dimensional analysis reveals significant variables. The effect of the length of the atomizer on SMD is found to be insignificant and the effect of other variables are also studied. An increase of the spiral angle reduces the exit orifice diameter and, thereby, decreases the Reynolds number which increases the spray quality. These results are consistent with the hydrodynamics of an atomizer. A correlation for SMD estimation is also proposed based on the experimental results.
2. The GA-based method for reverse analysis is used and the investigation of sets of solutions relies on human inspection. The inference procedure may include the selection of favourable SMDs based on different applications (propeller or thruster), flow conditions, or geometrical factors.

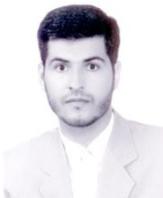
Maziar Shafaee received a B.S. degree in Mechanical Engineering from Tabriz University in 2000. He also received an M.S. and a Ph.D. degree from the University of Tehran in 2002 and 2011, respectively. Mr Shafaee is currently working as an assistant professor at the Faculty of New Sciences and Technology at the University of Tehran. His research interests include the numerical and experimental study of spray and atomization systems.

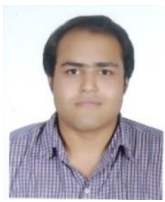
Armin Abdehkakha received a B.S. degree in Aerospace Engineering from the Sharif University of Technology in 2012. He has been a master's student at the University of Tehran since 2012. Mr Abdehkakha is currently working on his M.S. thesis at the Faculty of New Sciences and Technology at the University of Tehran.

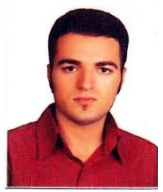
Abbas Elkaie received a B.S. degree in Mechanical Engineering from Babol University in 2012. He has been a master's student at the University of Tehran since 2012. Mr Elkaie is currently working on his M.S. thesis on the optimization of engineering systems at the Faculty of New Sciences and Technology at the University of Tehran.